\documentclass[runningheads]{llncs}

\usepackage[labelfont={small,bf},textfont={small,it},skip=1pt]{caption}
\usepackage{graphicx}
\usepackage{subcaption}
\usepackage{booktabs}
\usepackage{caption}
\usepackage{tabularx}
\usepackage[normalem]{ulem}
\usepackage{fancyref}
\usepackage{xcolor}
\usepackage{float}
\usepackage{cite}
\usepackage{amssymb}
\usepackage[skip=0.1cm]{parskip}

\graphicspath{{figures/}{./}}

\useunder{\uline}{\ul}{}

\begin{document}

\title{Measuring the Prevalence of WiFi Bottlenecks in Home Access Networks}
\author{Ranya Sharma, Marc Richardson, Guilherme Martins, Nick Feamster}
\institute{University of Chicago}

%%%%%%%%%%%%%%%%%%%%  START OF DOCUMENT  %%%%%%%%%%%%%%%%%

\maketitle

\begin{sloppypar}
\begin{abstract}
    As broadband Internet speeds continue to increase, the home wireless
    (``WiFi'') network may more frequently become a performance bottleneck.
    Past research, now nearly a decade old, initially documented this
    phenomenon through indirect inference techniques, noting the prevalence of
    WiFi bottlenecks but never directly measuring them. In the intervening
    years, access network (and WiFi) speeds have increased, warranting a
    re-appraisal of this important question, particularly with renewed private
    and federal investment in access network infrastructure. This paper
    studies this question, developing a new system and measurement technique
    to perform direct measurements of WiFi and access network performance,
    ultimately collecting and analyzing a first-of-its-kind dataset of more
    than 13,000 joint measurements of WiFi and access network throughputs, in
    a real-world deployment spanning more than 50 homes, for nearly two years.
    Using this dataset, we re-examine the question of whether, when, and to
    what extent a user's home wireless network may be a performance
    bottleneck, particularly relative to their access connection. We do so by
    {\em directly} and {\em continuously} measuring the user's Internet
    performance along two separate components of the Internet path---from a
    wireless client inside the home network to the wired point of access
    (e.g., the cable modem), and from the wired point of access to the user's
    ISP. Confirming and revising results from more than a decade ago, we find
    that a user's home wireless network is often the throughput bottleneck. In
    particular, for users with access links that exceed 800~Mbps, the user's
    home wireless network was the performance bottleneck 100\% of the time.
\end{abstract}

\section{Introduction}\label{sec:intro}

Home Internet users invest significant resources and expense improving the
state of their wired Internet access connections, which in the United States,
now routinely exceed 1~Gbps for many service plans.  As wireline broadband
access Internet speeds continue to increase, however, home Internet users are
increasingly faced with a new performance bottleneck: their home wireless
network. The prospect of the home wireless network as an end-to-end performance
bottleneck is, in fact, not new: nearly a decade ago, researchers noted that
WiFi networks in homes routinely introduced throughput bottlenecks on home
networks when the fixed-line ISP throughput exceeded 35~Mbps. Wireless
technologies have improved substantially over the last decade, naturally, yet
fixed-line Internet throughput {\em also} has continued to increase,
warranting a re-appraisal of this fundamental question.

The stakes for understanding home Internet performance bottlenecks has perhaps
never been higher, as federal funding is poised to invest in Internet
infrastructure across the United States.  Understanding this question is
critically important, especially as increasing attention turns to
understanding and mapping Internet speed across the United States and
globally. In many cases, Internet speed measurements are conducted with
crowdsourced Internet measurement tools (e.g., Ookla's SpeedTest, Measurement
Labs's NDT), and previous work has observed that client-based speed tests can
often be the bottleneck in the end-to-end performance.

Indeed, Internet infrastructure does not stop at the ``wall jack'', and the
user experience often depends as much on infrastructure {\em inside} the home
as it does on infrastructure {\em to} the home. It does little practical
good for an Internet service provider to deliver fast speeds over a wired
connection to a single point in the home if the wireless connectivity that
users rely on in the rest of the home is poor. Indeed, this concern is not a
hypothetical but a very real problem for many Internet users. In April 2021, a
Pew Research Center Survey that ``roughly half of [U.S.] broadband users
report[ed] that they often (12\%) or sometimes (37\%) experienced problems
with the speed, reliability or quality of their high-speed Internet
connections at home” and that ``29\% of broadband users did something to
improve the speed, reliability or quality of their [home Internet] since the
beginning of the pandemic''. Many people experience
problems with their home Internet performance and want to improve it. 

Nearly a decade ago, researchers showed that in legacy 802.11 networks, home wireless
connectivity was the performance bottleneck in many homes, particularly as
access speeds from ISPs exceeded 35~Mbps~\cite{sundaresan2016:pam}.  Since then, wireless technologies 
have improved substantially, and ISPs have continued to increase access
speeds.  In this paper, we revisit the question of whether home wireless
connectivity is a performance bottleneck in the modern era.  We find that
home wireless connectivity is still a performance bottleneck in many homes,
particularly for users whose ISP access speeds exceed 100~Mbps; for those
whose speeds exceed 800~Mbps, the wireless network is the bottleneck 100\% of
the time.

This paper makes the following contributions:
\begin{itemize}
\itemsep=-1pt
\item We develop the first method to {\em jointly} and {\em continuously} measure the
  performance of the wired and wireless networks in a home network, using a
        combination of (1)~a router-based speedtest performing measurements over
        the wide-area from the user's wired network access point; (2)~a
        browser-based plug-in that performs continual throughput measurements
        from a user's wireless device inside the home to a speedtest server
        running inside the home network at the router.
\item We affirm and amend the presence of wireless performance bottlenecks in
    the home network, finding that the prevalence of WiFi bottlenecks
        increases significantly for homes with access speeds exceeding
        100~Mbps, and that homes whose ISP speed tiers exceed 800 Mbps, the
        home WiFi network was the bottleneck in nearly 100\% of the
        measurements we performed.
\item We publicly release our measurement toolkit---router-based speedtest,
    browser-based home network speedtest, data, and associated analysis---to
        encourage the continued reassessment of these results, in a broader
        range of settings, and over time as ISP access networks and WiFi technologies 
        continue to evolve.
\end{itemize}

\section{Background and Related Work}\label{sec:background}

In this section, we provide background on WiFi measurements in access
networks, as well as previous work that has performed measurements of WiFi in
different capacities and contexts.

\subsection{Background}

\if 0
\begin{figure}[t!]
  \includegraphics[width=\columnwidth]{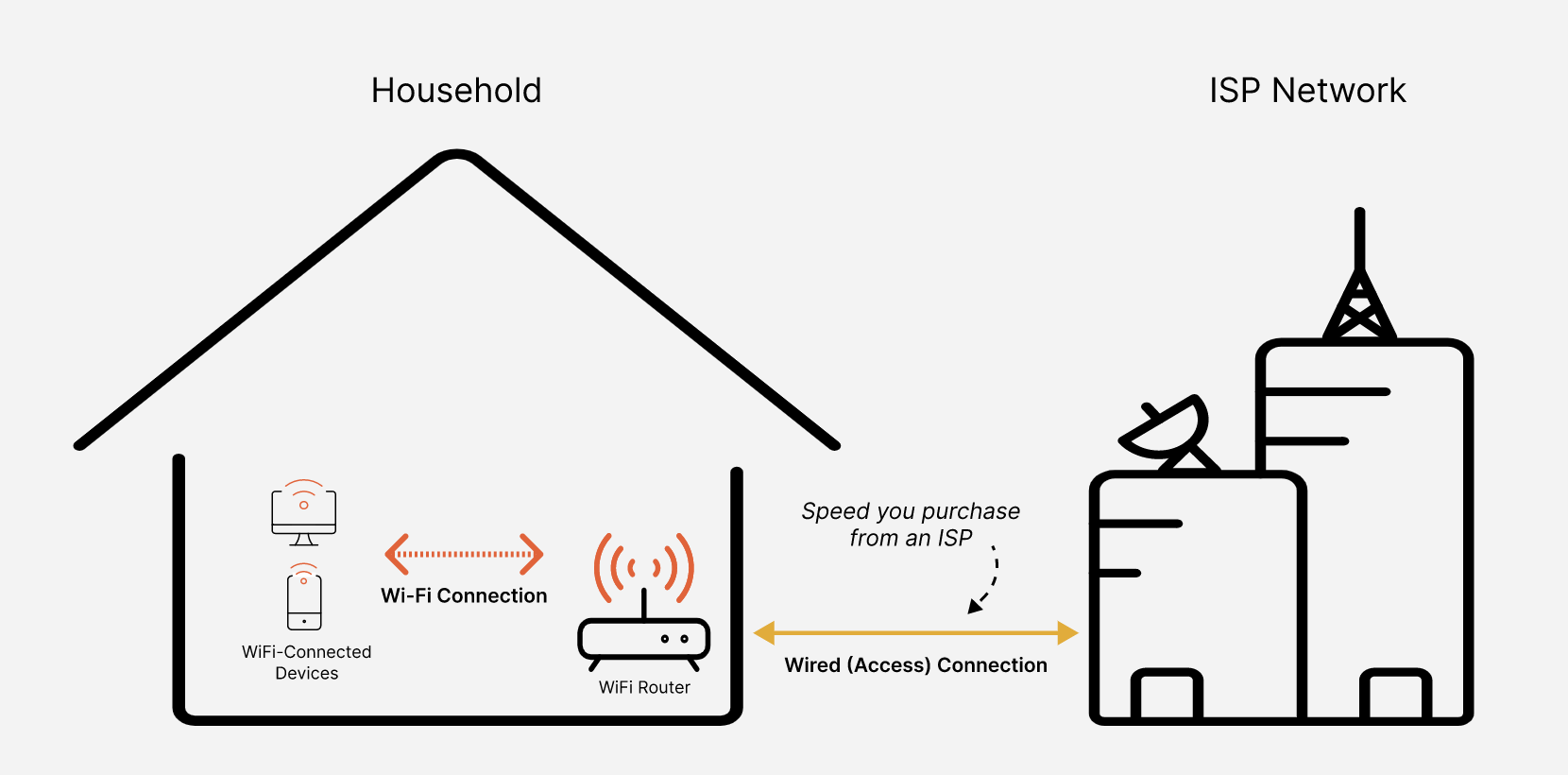}
  \caption{WiFi Connection and Wired (Access) Connection in a Household.}
  \label{fig:wifi_household}
\end{figure}
\fi

Wireless Internet access via the 802.11 protocol (``WiFi'') has become
increasingly prevalent in recent years.  As of 2021, 93\% of households in the
United States had an Internet connection, and the vast majority of these
households rely on WiFi for connectivity within the home~\cite{pew-broadband}.
The number of devices connected to the Internet has also increased
dramatically.  As of 2020, the average U.S. household has 25
Internet-connected devices, more than double the number of connected devices
in 2019~\cite{deloitte-devices}, likely in large part due to the proliferation of
Internet of Things (IoT) devices in homes.  As wireless Internet connectivity
in homes becomes increasingly prevalent, it has become increasingly important
to understand whether, and to what extent, home wireless Internet connectivity
affects the performance that users ultimately experience in their homes.

A common mode of connectivity in a home network is to first connect to a WiFi
router in an Internet user's home, which then connects the user through a wired network to the
rest of the Internet. A household’s WiFi connection typically has different
performance characteristics than the corresponding wired connection. WiFi
connection performance is determined by factors independent of your access
connection, while the access connection performance is primarily determined by
the specific service contract you purchase from a service provider (ISP). WiFi
bottlenecks occur when the speed of your WiFi connection is worse than the
speed of the access connection.

\begin{figure}[t!]
    \centering
    \begin{subfigure}[b]{0.75\columnwidth}
        \includegraphics[width=\columnwidth]{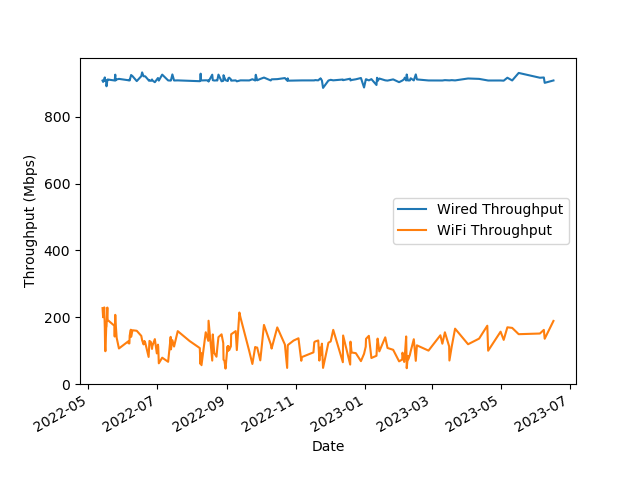}
        \caption{WiFi bottleneck.}
        \label{fig:wifi_bottleneck}
    \end{subfigure}
    \begin{subfigure}[b]{0.75\columnwidth}
        \includegraphics[width=\columnwidth]{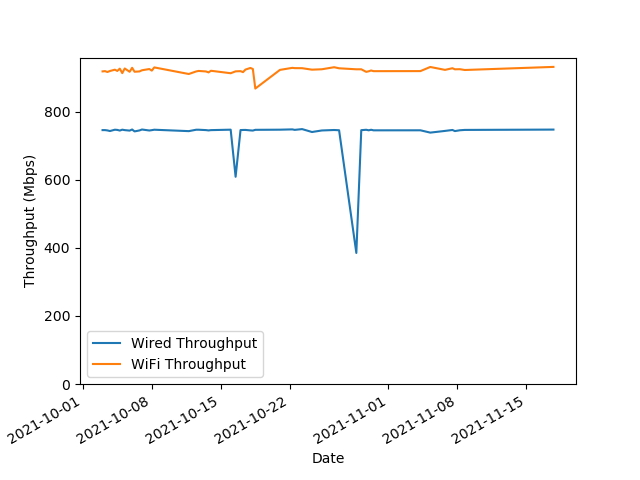}
        \caption{Access bottleneck.}
        \label{fig:access_bottleneck}
    \end{subfigure}
  \caption{WiFi and wired bottlenecks in households in a large city in the
    United States over the past 21 months.}
  \label{fig:household_bottleneck}
\end{figure}

Figure~\ref{fig:household_bottleneck} shows WiFi and access speeds from for
two households in a large city in the United States. Figure~\ref{fig:wifi_bottleneck}
has WiFi speeds below wired access speeds.  WiFi is a bottleneck 100 percent
of the time for this household. Conversely, in
Figure~\ref{fig:access_bottleneck},
WiFi speed exceeds access speed. Note that the household with WiFi bottlenecks is paying
for Internet service with high access speeds (about 900~Mbps) whereas the
household with no bottlenecks is paying for slightly lower access speeds (some
of the households in our study have slower access speeds than this example).
As the results from our paper show, the prevalence of WiFi bottlenecks is
highly correlated with the access speed that a household purchases.

WiFi bottlenecks can significantly affect a user's online experience.  For
example, the access ISP connection might support high speeds, but if the
corresponding WiFi connection is poor, a user will ultimately experience the
network speeds corresponding to the slower WiFi connection.  WiFi bottlenecks
also have important implications for policies aiming to improve Internet
performance in a community. For example, significant attention (and funding)
is currently being devoted to improving Internet infrastructure. Yet,
ultimately, if performance bottlenecks are in home network
infrastructure---and in particular the home wireless network---then
investments must also be directed towards assessing and improving the WiFi
equipment that households use to access the Internet.

\subsection{Related Work}

Various previous work has attempted to characterize wireless network
performance in specific settings, including home networks. Early seminal work
from Sundaresan et al.~\cite{sundaresan2016:pam} characterized the performance of
802.11b networks in a home setting, and found that when access link speeds
exceeded approximately 30 Mbps, the home wireless network became the
bottleneck link. This study's measurement techniques were
indirect: the technique used inferred the presence
of bottlenecks based on inter-packet arrivals downstream of the wireless
access point. In contrast, this work measures wireless network performance
{\em directly}. Additionally, this previous work was conducted nearly a decade ago;
this work thus constitutes a re-appraisal of this past work.

Previous work from Shi et al.~\cite{shi:walk-on-client-side} and Chung et
al.~\cite{chung:measurements-mobile-traffic} have also collected client-side
wireless network measurements. However, these studies do compare wireless
measurements to wired measurements, as ours does. Instead,  Chung et al.
present the relationship between application traffic and its origin device.
Shi et al. strictly collect channel scans, presenting the value of smartphone
channel scans.  Sui et al.~\cite{sui:characterizing-improving-wiFi-latency}
also design and deploy a framework for collecting WiFi measurements. However,
their framework measures and characterizes WiFi latency, while ours studies
WiFi throughput. 

More recently, Feamster and Livingood highlighted the challenges with
performing client-based speed tests~\cite{feamster2020:cacm:speed}, and
presented evidence that in many cases, home wireless access links are the
bottleneck in end-to-end Internet speed tests. This work, however, was 
a survey of the shortcomings of client-based speed tests, of
which wireless bottlenecks are one example. This work focuses
specifically on the prevalence of wireless bottlenecks in home networks.

Many other studies have studied the prevalance of wireless performance in
other contexts, although they have not explicitly focused on the extent to
which poor WiFi performance introduces a bottleneck along and end-to-end
path.  
Yang et al. developed a machine learning-based approach for
predicting performance degradation in WiFi networks, showcasing the potential
of predictive analytics in managing network performance~\cite{yang2021ml}.
In 2016, Sui et al. investigated the impact of interference on WiFi
performance, emphasizing the significance of interference detection and
mitigation strategies~\cite{sui2016understanding}. Sui et al. also developed
mechanisms for characterising and improving WiFi latency in large-scale
operational networks~\cite{sui2016characterizing}. 
Other work
has explored WiFi performance in high-density
environments~\cite{maity2016tcp}, while other work has explored the
implications of security mechanisms on WiFi performance, shedding light on how
encryption protocols can contribute to bottlenecks~\cite{lepaja2018impact}.
In the context of the Internet of Things (IoT), Sheth et al. investigated how
device heterogeneity contributed to bottlenecks in WiFi networks
\cite{sheth2019enhancing}. 
Other work has explored the effects of Quality of Service (QoS) mechanisms on
WiFi bottleneck identification~\cite{vollero2004frame}. 
Recent work from Grazia explored the impact of emerging technologies, such as Wi-Fi 6, on
TCP bottlenecks \cite{grazia2021future}.

\section{Method}\label{sec:method} 

\paragraph{Measurements: Netrics} We used the public, open-source Netrics
platform~\cite{netrics} to conduct periodic downstream throughput measurements
for both the access ISP connection and for the internal WiFi connectivity.
Netrics is a widely accepted measurement platform, now in use across the
United States for many studies of access Internet performance, that has been
underway for nearly two years.  It currently provides many open-source network
measurement tests, including a variety of continuous throughput and latency
tests.  A common way of performing Netrics measurements is by distributing a
single-board computer (typically, a Raspberry Pi) to study participants at
vantage points of interest.  The participants connect the Pis directly to
routers or modems in their home networks, and the Pis measure and archive
these performance metrics. The Netrics software distribution can schedule any
network measurement and is commonly deployed to run three different periodic
throughput tests: Measurement Labs's NDT7, Ookla's Speedtest, and iperf3 to a
nearby server. The measurement suite also runs a variety of other measurements
(e.g., DNS lookup latency, packet loss rates) that we do not analyze.

\begin{figure}[t!]
  \includegraphics[width=\columnwidth]{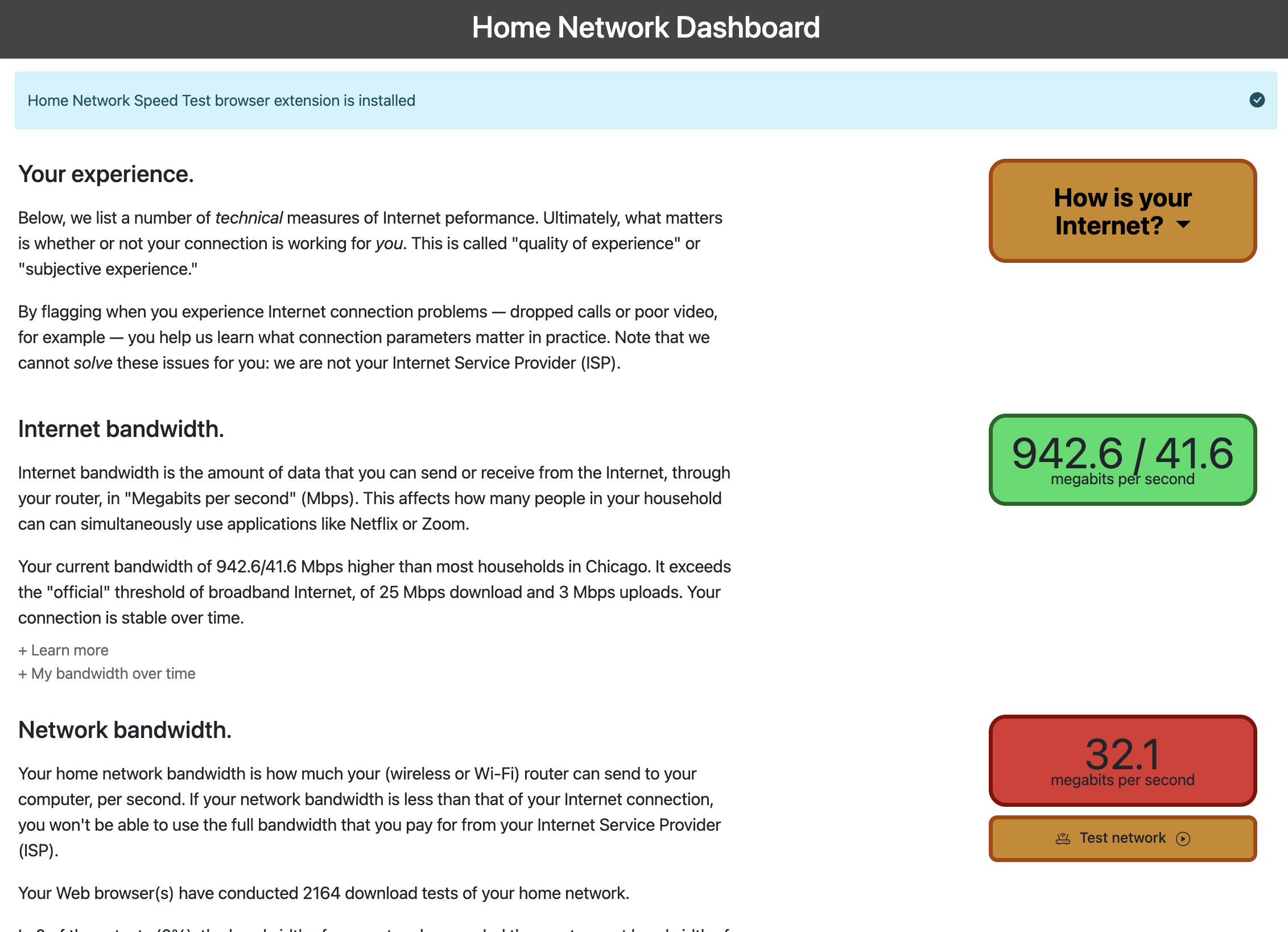}
    \caption{Netrics dashboard at \url{https://netrics.local/}.}
    \label{fig:plugin}
\end{figure}

\paragraph{Contemporaneous measurements of WiFi and ISP.} Directly comparing
the throughput of the (typically wired) access ISP network to that of the WiFi
network from the same household is extremely challenging.  It requires
deploying infrastructure in the home at the router both to conduct throughput
measurements of the access ISP and to host a server for conducting wireless
throughput measurements; it requires deploying software on one or more client
devices within the home network to measure the home WiFi; and, finally, it
requires collecting contemporaneous measurements of both the access ISP
throughput and the home wireless network. To perform the throughput
measurements from the home wireless network, we hosted a speed test server on
a Raspberry Pi, intended to be deployed at or near the home router. A user
could then direct his or her browser to the server at
\url{https://netrics.local/}, which would trigger the browser to run a
Javascript-based NDT7 throughput test to the server running on the Raspberry
Pi. To enable {\em continuous} measurements, we also implemented a public,
open-source Chrome browser plugin that would run periodic NDT7 tests from the
user's browser to the server running on the Raspberry Pi, even if the user
didn't explicitly initiate a test. We designed these throughput tests to run
every few hours, but the actual frequency of the tests depended on whether
someone was using device on which the browser plugin was installed.
Figure~\ref{fig:plugin} shows an example of the public, open-source Netrics
dashboard, which displays the results of the contemporaneous measurements of
the access ISP and the home WiFi network. This particular example shows a case
where the measured ISP access throughput (shown in green) is far greater than
the measured WiFi throughput (shown in red), indicating that WiFi is the
bottleneck.

\paragraph{Participant recruitment.} We installed Netrics on a fleet of
Raspberry Pis and recruited participants through community-based organizations
and social media channels.  Participants were offered a monetary incentive of
\$100 to install a device to an access point for one month, an additional \$25
to continue hosting the device for another six months, and no incentives after
that (akin to a ``RIPE Atlas model'' of ongoing deployment). We have been
collecting data from the deployment since June 2021 among our research team,
and since October 2021 across a broader set of study participants across one
of the largest cities in the United States. 

\begin{table}[t]
    \begin{center}
    \begin{tabular}{lr}
\\ \hline
        {\bf Period} & September 18, 2021 - June 30, 2023 \\
        {\bf Households} & 52 \\ \hline
        {\bf Vantage Points} & 65 \\
        - < 50 & 3 \\
        - 50-100 & 15 \\
        - 100-200  & 3 \\
        - 200-400  & 10 \\
        - 400-800  & 21 \\
        - > 800 Mbps & 13 \\
        \hline
        {\bf Measurements} & 13,581 \\ \hline
    \end{tabular}
    \end{center}
    \caption{Summary of the data collected from the Netrics deployment.}
    \label{tab:measurements}
\end{table}

\paragraph{Data collection.} Table~\ref{tab:measurements} summarizes the
measurements for the study. The measurements collected for this study were
collected from September 18, 2021 through June 30, 2023.  To measure download
throughputs of both the access link (the access speed) and the wireless
network (the WiFi speed), we use Measurement Lab’s Network Diagnostic Tool
(NDT7) tool.  We performed access throughput measurements for each household
approximately once per hour.  WiFi speed measurements did not follow a fixed
schedule, due to the fact that the WiFi measurements were taken from a user's
browser on a wireless device in the home and thus depended on the user
actively using the device where the browser plug-in was installed.  To ensure
that the measurements compared were taken at around the same time, we
re-sampled the data for each household.  We used a 6-hour time window and
aggregated using the median access and WiFi speed within each window. 
We used a 6-hour time window because we found that this was the time period in which there
was the most samples. This time window had a broad enough window to contain 
a large amount of data, yet small enough intervals to sufficiently splice the data. 

\begin{figure}[t!]
  \includegraphics[width=\columnwidth]{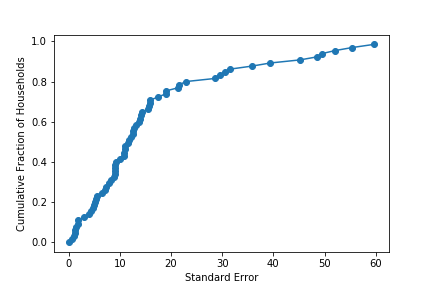}
  \caption{Sample Error for the Difference in WiFi and NDT Values per Household}
  \label{fig:standard_error}
\end{figure}

\paragraph{Data analysis.} We restricted our sample of households to those that
had a significant number of WiFi measurements during the analysis period,
dropping households that did not have an enough WiFi measurements to allow us
to draw strong conclusions about the performance of the wireless network and
corresponding wired access network.  After resampling the data for each
household, we dropped households from the analysis that had less than 20
measurements with both WiFi speed and access speed. To determine this cutoff, we found the sample error for the difference between WiFi and Internet speed measurements per hosuehold. Figure~\ref{fig:standard_error} is the CDF of sample error that we created based on these findings, through which we found that the cutoff must be 20 measurements. After dropping households from the dataset, our sample had 52 households, with an average of
261 WiFi and access speed coincident wireless and wired measurements per household.

Some households experienced a significant change in access speeds during the
period of analysis, likely due to a change in Internet service provider or
service plan.  This occurred for 13 households in our sample.  Because these
changes in access speeds may significantly affect our analysis (i.e., by
changing the wired throughput bottleneck for a household), we split the data for these
households and treat each as a separate observation point.  This ensured that the
access speed is constant during the analysis period.  As a result, we add
13 observation points to our sample, bringing the total number of observation
points to 65.

\paragraph{Limitations.} In our study, we lack insight into the types of devices
that study participants use. Furthermore, we do not have complete visibility into the
environment that the measurements are taken in. This does not affect our overall 
conclusions but bounds the types of conclusions that can be made. Our conclusions 
are still valid in a broader context--we identify that bottlenecks exist but do not 
know what they are directly attributed to. We are unable to draw conclusions about 
whether the bottleneck is generated by properties of the network or hardware.
Additionally, our study primarily takes place in the Chicago area. However, we do not 
consider this to be a limitation that affects our conclusions because characteristics 
of WiFi are unlikely to vary by geography. 
\section{Results}\label{sec:results} 

To evaluate performance bottlenecks at the household level, we examine throughput
test data from the Netrics devices across the deployment for the 21-month
period of the study.  After measuring the download throughputs of the WiFi and
access connections, we compare the WiFi throughput of each household to its
downstream access throughput over time, paying particular attention to the
throughput of the WiFi network relative to corresponding ISP access network.

\subsection{Prevalence of WiFi Bottlenecks}\label{sec:common_bottlenecks}

\begin{figure}[t!]
    \centering
  \includegraphics[width=0.8\columnwidth]{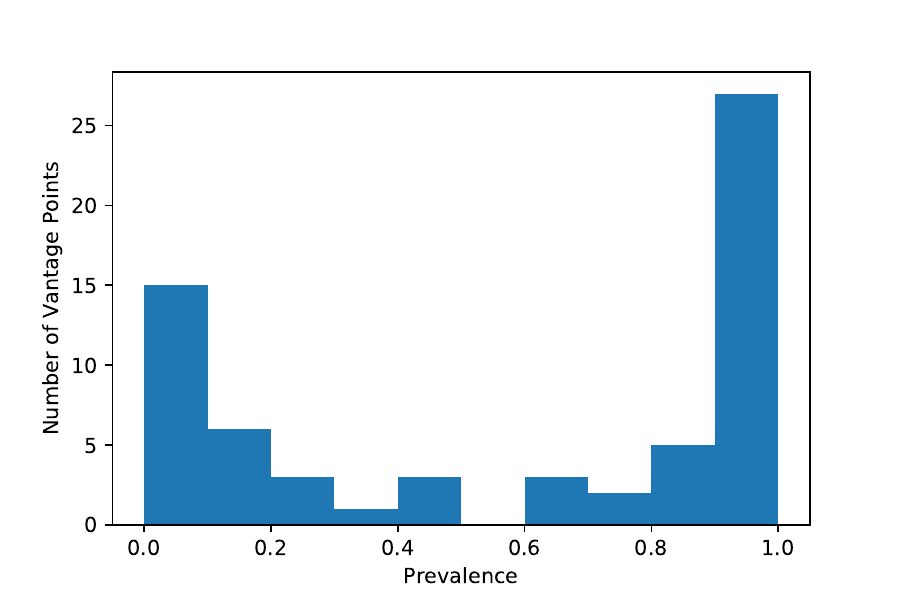}
  \caption{Prevalence of WiFi bottlenecks by vantage point.}
  \label{fig:bottleneck_prevalence}
\end{figure}

Figure~\ref{fig:bottleneck_prevalence} highlights the prevalence of WiFi
bottlenecks across households in our study.  To calculate the prevalence for each
household, we compare the household’s access throughput to its WiFi speed over
time and note extent to which the WiFi throughput is lower than the access
throughput.  For example, if a household has 100 measurements of WiFi and
access throughputs and, in 20 of 100 measurements, WiFi throughput is below
access speed, then the bottleneck prevalence for this household would be 0.2;
89\% of households experienced at least one WiFi
bottleneck.
Across vantage points, the vast majority either experience no WiFi bottlenecks
or a high percentage, as Figure~\ref{fig:bottleneck_prevalence} shows. Most
vantage points do not have a prevalence between 10\% and 80\%, indicating 
in most cases, either WiFi is almost never a bottleneck, or it is almost
always a bottleneck. As we will see, the circumstance depends on the
household's access speed tier.

\begin{figure}[t!]
    \centering
  \includegraphics[width=0.8\columnwidth]{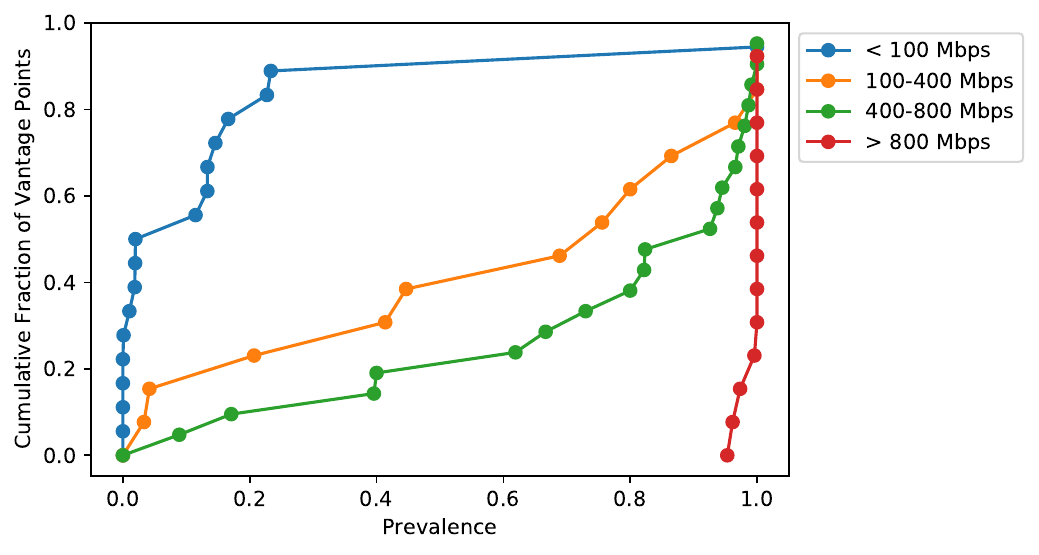}
  \caption{Prevalence of WiFi bottlenecks by speed tier.}
  \label{fig:bottleneck_prevalence_tier}
\end{figure}

Figure~\ref{fig:bottleneck_prevalence_tier} shows the prevalence of WiFi
bottlenecks across vantage points, according to the speed tier of the access network.
Each point on each curve represents a vantage point; the x-axis represents the
prevalence of WiFi bottlenecks at that vantage point, and the y-axis
represents is the cumulative distribution of vantage points.  As expected,
vantage points with higher speed tiers see a greater prevalence of WiFi
bottlenecks. For vantage points with access speed tiers of less than 100 Mbps,
the prevalence of WiFi bottlenecks, tends to be less, with the median vantage
point seeing a prevalence of less than 0.1. On the other hand, for vantage
points whose speed tier exceeds 800 Mbps, the prevalence of WiFi bottlenecks
is 1, indicating that {\em all} joint measurements in those households
indicate the presence of a WiFi bottleneck.  

% \begin{figure}[t!]
%   \includegraphics[width=\columnwidth]{bottleneck_prevalence_access_throughput}
%   \caption{WiFi bottlenecks are more prevalent with higher access throughput.}
%   \label{fig:bottleneck_access_throughput}
% \end{figure}

This variation of the prevalence of WiFi bottlenecks also correlates with the
measured throughput of the access ISP.  The median access ISP throughput among
households that experience frequent bottlenecks is 700.52~Mbps; on the other
hand, the median access ISP throughput is only 58.12 Mbps for households that
rarely experience bottlenecks. We find that WiFi bottlenecks become increasingly prevalent as measured access
throughputs increase.

% Figure~\ref{fig:bottleneck_access_throughput}
% shows all measurements of WiFi and access speeds for all households in the
% study.  A red point indicates that the WiFi throughput was above the access
% speed at the time of measurement.  A yellow point indicates that the WiFi
% throughput was below the access speed at the time of measurement.  The figure
% shows that WiFi bottlenecks become increasingly prevalent as measured access
% throughputs increase. 

\subsection{Magnitude of WiFi Bottlenecks}\label{sec:wifi_compare_access}

% \begin{figure}[t!]
%   \includegraphics[width=\columnwidth]{speed_tier_prevalence}
%   \caption{Prevalence of WiFi bottlenecks for measurements in each speed tier.
%     Measurements on broadband access networks with higher speed tiers show a
%     greater prevalence of WiFi bottlenecks.}
%   \label{fig:plot_by_throughput_tier}
% \end{figure}

To evaluate the magnitude of a WiFi bottleneck (i.e., how much slower the WiFi
link is than the ISP access link), we compare the median WiFi throughput,
access throughput, and the effective throughput of a household, which we
define as the minimum of the median WiFi throughput and the median access ISP
throughput.  

% Figure~\ref{fig:plot_by_throughput_tier} plots the median access
% throughput, actual throughput, and WiFi throughput across all measurements for
% all households within each speed tier.

If WiFi throughput is similar to or greater than access ISP throughput, then
its value should consistently be greater than its corresponding access ISP
throughput value.  In this case, because the access throughput is the lower
value, the median actual speed should approximately equal the median access
speed for each speed tier.  This occurs for the first three throughput tiers,
shown with actual speed and access speed having very close values.  However,
the values begin to diverge for speed tiers that exceed 200~Mbps.  As
expected, access ISP throughputs increase as ISP speed tier increases.
However, effective throughputs plateau, as the household's WiFi network
becomes the bottleneck at higher ISP throughputs (and speed tiers).

The extent of the difference between the actual throughput and the access
speed for each tier indicates the magnitude of the WiFi bottlenecks.  For
example, we find that households with
speed tiers in the 200--400 Mbps range experience a mean actual throughput
of 155.69~Mbps, while the mean access throughput is 265.60~Mbps.  This disparity
shows that, while paying for access throughputs between 200 and 400 Mbps, the
median household is typically experiencing actual speeds that are 109.91 Mbps
lower.  Interestingly, this gap increases with higher throughput tiers.
For households in the 400--800 Mbps tier, the gap between access throughput and
actual speed grows to 240~Mbps, and for households in the 800 Mbps plus tier
to 602 Mbps.  This result indicates that home wireless networks introduce
bottlenecks that often result in households failing to completely utilize the
access throughput that they have purchased from their ISP.

\section{Discussion}\label{sec:discussion}

The results we have presented in this paper are, by many accounts,
unsurprising, as both decades-old research and continual anecdotes have
suggested that WiFi is often the bottleneck in home networks. Nonetheless,
such a study remains important, for several reasons. 

First, the contribution of the measurement method and system itself, the
first system to continuously and contemporaneously measure home WiFi
throughput and access network throughput from the same vantage point over an
extended period of time.  We have already released these tools publicly, which
are currently in use by state broadband offices across the United States, as
well as many other community-based organizations as they try to understand the
extent of performance bottlenecks in access networks.

Second, the implications of these findings---as well as those that will come
from continued and future measurements of this phenomenon from our
software---have important implications for investment in future broadband
infrastructure. In the United States, for example, the Broadband Equity,
Access, and Deployment (BEAD) program, has currently allocated more than \$40 billion
to states to upgrade their broadband infrastructure. An important key to
spending these funds wisely involves knowing precisely where performance
bottlenecks lie, and what types of upgrades will ultimately affect the lived
experience of Internet users. 

For state and local broadband offices, policies to improve online experience
in a community must reckon with the prevalence of WiFi bottlenecks and their
effect on lived Internet performance. Although investments in the last-mile
and middle-mile infrastructure are important, they will not significantly
improve a community’s online experience if they are not also supplemented with
policies to improve the equipment that people use to access the Internet
within their homes, such as their WiFi equipment. Community organizations, and
state and municipal broadband offices should play close attention to
improving WiFi networks in communities.

Our results also have important implications for consumer protection. If a
user's home WiFi network is a bottleneck along the end-to-end path, then
upgrading to a higher service plan may not necessarily improve a home user's
Internet experience.  Our results demonstrate that WiFi is a persistent
bottleneck in homes whose access ISP throughput exceeds 200~Mbps.  Consumers
in these categories may thus be better served by first improving their home
WiFi networks before upgrading their service plans.

\label{lastpage}\section{Conclusion}\label{sec:conclusion}

This paper presents a first-of-its kind study of {\em direct, coincident}
measurements of access throughput and WiFi throughput in the same home
broadband access network. It is also, to our knowledge, the first such study
in nearly a decade, hence constituting a much-needed reappraisal of the
extent to which WiFi is a bottleneck in home broadband access networks. 

One should not expect to be surprised by our results. The results of this
study are completely expected, especially in light of past research in this
area that has highlighted the presence of WiFi bottlenecks in home
networks~\cite{sundaresan2016:pam}. The purpose of this paper is not to
present a new, ``surprising'' result, but rather to shed light and provide
concrete evidence from home networks in deployment to demonstrate the
prevalence of these bottlenecks in practice.  Although the state of home WiFi
is known anecdotally, for many audiences, especially for ISPs, operators, and
regulators, it is important to have both documentation and public data
confirming the existence of these bottlenecks in real networks, as well as a
framework for repeating these measurements in other networks in the future.
This paper presents all of these contributions. These results are particularly
important to document in light of ongoing investments in Internet
infrastructure to bring faster speeds to homes via access ISPs. 

\pagebreak
\bibliographystyle{splncs04}
\bibliography{paper, ref}

\end{sloppypar}
\end{document}